\documentclass[]{raa}
\usepackage{graphicx,times}
\usepackage{natbib}
\usepackage{amssymb,amsmath}
\usepackage{rotating}
\usepackage{longtable}
\usepackage{supertabular}
\usepackage{longtable,lscape}
\usepackage{float}
\usepackage{natbib}
\usepackage{amsmath}
\usepackage{amssymb}
\usepackage{graphicx}
\usepackage{xspace}
\bibpunct{(}{)}{;}{a}{}{,}

\def\apj{ApJ}
\def\aap{A\&A}
\def\mnras{MNRAS}
\def\apjs{ApJS}
\def\apjl{ApJL}

\begin{document}

   \title{The core dominance parameter and \emph{Fermi} detection of extragalactic radio sources
$^*$
\footnotetext{\small $*$ }
}

 \volnopage{ {\bf 201?} Vol.\ {\bf X} No. {\bf XX}, 000--000}
   \setcounter{page}{1}

   \author{Zhenkuo Liu\inst{1}, Zhongzu Wu\inst{1}, Minfeng Gu
      \inst{2}
   }

   \institute{ College of Science, Guizhou University, Guiyang 550025, China.; {\it zzwu08@gmail.com}\\
        \and
              Key Laboratory for Research in Galaxies and Cosmology, Shanghai Astronomical Observatory, Chinese Academy of
Sciences, 80 Nandan Road, Shanghai 200030, China\\
\vs \no
   {\small Received ; accepted }
}

\abstract{
In this paper, by cross-correlating an archive sample of 542 extragalactic radio sources with the \emph{Fermi}-LAT Third Source Catalog(3FGL), we have compiled a sample of 80 $\gamma$-ray sources and 462 non-\emph{Fermi} sources with available core dominance parameter($R_{CD}$), core and extended radio luminosity; all the parameters are directly measured or derived from available data in the literature. We found that $R_{CD}$ have significant correlations with radio core luminosities, $\gamma$-ray luminosity and $\gamma$-ray flux respectively; the \emph{Fermi} sources have on average higher $R_{CD}$ than non-\emph{Fermi} sources. These results indicate that the \emph{Fermi} sources should be more compact, and beaming effect should play a crucial role for the detection of $\gamma$-ray emission.  Moreover, our results also show \emph{Fermi} sources have systematically larger radio flux than non-\emph{Fermi} sources at fixed $R_{CD}$, indicating larger intrinsic radio flux in \emph{Fermi} sources. These results show a strong connection between radio and $\gamma$-ray flux for the present sample and indicate that the non-\emph{Fermi} sources is likely due to low beaming effect, and/or the low intrinsic $\gamma$-ray flux, support a scenario in the literature: a co-spatial origin of the activity for the radio and $\gamma$-ray emission, suggesting that the origin of the seed photons for the high-energy $\gamma$-ray emission is within the jet.
}

   \authorrunning{Liu et al. }            
   \titlerunning{}  
   \maketitle

%
\section{Introduction}           
\label{sect:intro}

Blazars are the most extreme active galactic nuclei(AGN) with characteristic properties such as large and variable polarization, apparent superluminal motion, flat or inverted radio spectra, and a broad continuum from radio through $\gamma$-rays \citep[e.g.,][]{urry95}. Because of the launch of the \emph{Fermi} satellite, the whole $\gamma$-ray sky almost has been scanning once every three hours since July of 2008 by the Large Area Telescope \citep[e.g.,][]{atwood09} on board. The third LAT AGN catalog(3LAC)\citep{ackermann15} and \emph{Fermi}-LAT Third Source Catalog(3FGL)\citep{2015ApJS..218...23A} showed that among all the \emph{Fermi} detected  AGNs(FAGNs), nearly all of them are blazars, while it should be noted that there are far larger number of blazars and other type of AGNs that is not detected by \emph{Fermi}.

The differences between FAGNs and non-\emph{Fermi} AGNs(NFAGNs) have been addressed in the literature. \cite{piner12} showed that sources detected with \emph{Fermi} have higher apparent speeds than those sources not detected with \emph{Fermi}. \cite{pushkarev12} found that the FAGNs have higher brightness temperature and VLBI core flux densities. \cite{linford12} showed that the Fermi detected BL Lacs(FBLs) have longer jets and polarized more often. \cite{wu14} selected a sample of 100 FBLs and 70 non-\emph{Fermi} BL Lacs(NFBLs) and found that the Doppler factor and intrinsic radio flux is on average larger in FBLs than in NFBLs. Based on a large sample of blazars, \cite{xiong15} found that there are significant differences between \emph{Fermi} blazars and non-\emph{Fermi} blazars for black hole mass, jet kinetic power from "cavity" power, and the broad-line luminosity.

By now, Doppler boosting is believed to be one important answer for the question "why are some sources $\gamma$-ray loud and others are $\gamma$-ray quiet ?"\citep{2015ApJ...810L...9L,wu14,2011ApJ...726...16L}.
Doppler factor $\delta$ can directly measure the significance of jet beaming effect;  a reliable determination of the Doppler factor, $\delta$, is a key step in studying the physical process for the compact emission regions of AGNs \citep[e.g.,][]{wu07}. However, the Doppler factor calculation is quite difficult and no reliable method for all the sources\citep[e.g.,][]{wu07}. According to the beaming model of AGN,  the emissions are composed of two parts, that is, the boosted core and the isotropic extended ones \citep[e.g.,][]{fan03}. The $R_{CD}$ is calculated by using the ratio of two parts, $R_{CD}$=F$_{C}$/F$_{E}$, where F$_{C}$ and F$_{E}$ is the flux of the boosted core and extended structure respectively\citep[e.g.,][]{orr82}. On account of the jet emissions are very strong beamed, the $R_{CD}$ should present the orientation of the jet \citep[e.g.,][]{fan03}. To some extent, the $R_{CD}$ is associated with the beaming effect in AGNs \citep[e.g.,][]{fan11}. \cite{fan06} found a significant correlation between the $R_{CD}$ and the Doppler factor $\delta$ derived from the lowest $\gamma$-ray flux. It will be effective for us to use $R_{CD}$ instead of Doppler factor to understand the relation between beaming and $\gamma$-ray detection of our sources in this work.

Although it is believed that beaming is an important parameter for the detection of $\gamma$-ray flux, while other parameters are still unclear. \cite{wu14} have shown that Doppler factor is an important parameter of
$\gamma$-ray detection, the non-detection of $\gamma$-ray emission in BL Lacs is likely due to low beaming effect, and/or low intrinsic $\gamma$-ray
flux.  The one important aim of this paper is to test that if it is still valid for the $\gamma$-ray detection of other types of AGN.
This paper is organized as follows: the sample selection is stated in section 2; the results are showed in section 3; the discussion is presented in section 4, and the summary is given in section 5. Throughout the paper we define the spectral index $\alpha$ as $\rm
f_{\nu}\propto\nu^{-\alpha}$ where $f_{\nu}$ is the flux density at frequency $\nu$ and a cosmology with $\rm H_{0}=70
\rm {~km ~s^ {-1}~Mpc^{-1}}$, $\rm \Omega_{M}=0.3$, $\rm
\Omega_{\Lambda} = 0.7$. All values of luminosity used in this
paper are calculated with our adopted cosmological parameters.

\section{The sample}
\cite{fan03} present a large sample of 542 extragalactic radio sources(27
BL Lacs, 215 quasars, and 300 galaxies) with $R_{CD}$ at 5 GHz and other parameters. Under the assumption that the core spectral index is $\alpha_{C}$ = 0.0  and the extended spectral index is $\alpha_{E}$ = 0.5 and 1.0 respectively,  the $R_{CD}$is derived as: $R_{CD}$=$\frac{L_{C}}{L_{E}}$=$\frac{L_{C}}{L_{T}-L_{C}}$(1.4/5)$^{-\alpha_{E}}$(1+z)$^{-\alpha_{E}}$, where $L_{C}$ is the 5 GHz radio core luminosity, $L_{E}$ is the 5 GHz extended luminosity, and $L_{T}$ is the 1.4 GHz total luminosity(see \cite{fan03} for detail).

In this work,
we cross-correlate this sample with the 3FGL\citep{2015ApJS..218...23A}, this
offers a sample of 80 $\gamma$
-ray detected AGNs, including 22 BL Lac objects, 11 radio galaxies, 3 seyfert galaxies and 44 Quasars( see Table. 1).
In this work, the $\gamma$
-ray flux and luminosity is from 100 MeV to 300 GeV energy range. The corresponding results are listed in Table 1, of which Col.1 is the source name,
Col.2 is identification(BL stands for BL lacertae objects, Q for quasars, G for radio galaxy, S, S1,
S2 for Seyfert galaxies), Col.3 is redshift, Col.4 for total luminosity at 1.4 GHz ,
Col.5 for core luminosity at 5 GHz , Col.6 and Col.7 are $R_{CD}$
corresponding to $\alpha_{E}$
=0.5, 1.0 respectively, and Col.8 is the $\gamma$-ray luminosity.
In total, we have a sample of 542 extragalactic sources containing 80 \emph{Fermi} objects(22 BL Lacs, 44
quasars, and 14 galaxies) and 462 non-\emph{Fermi} objects(5 BL Lacs, 171 quasars, and 286 galaxies).

\section{The results}

To study the differences of FAGNs and NFAGNs,  we compare
various radio properties for two subsamples, including $R_{CD}$, the core and extended luminosity. The results are shown as follows.

\subsection{The distributions of $R_{CD}$ for FAGNs and NFAGNs}

\begin{figure*}
\centering
\includegraphics[width=8cm, height=6cm]{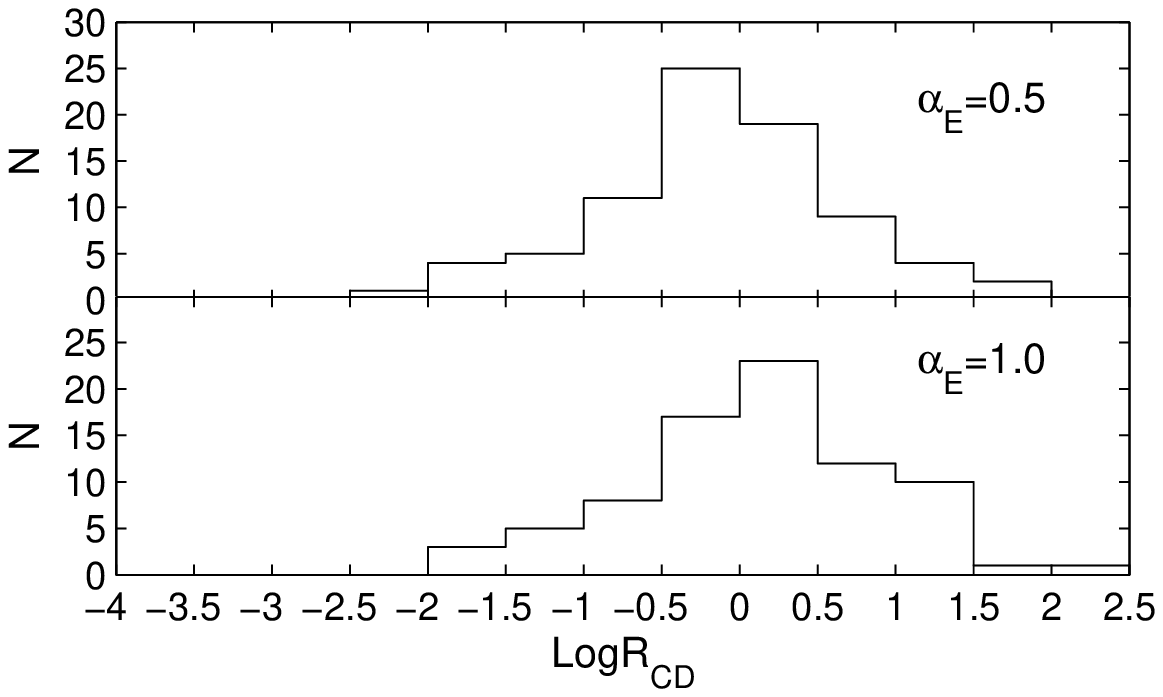}
\includegraphics[width=8cm, height=6cm]{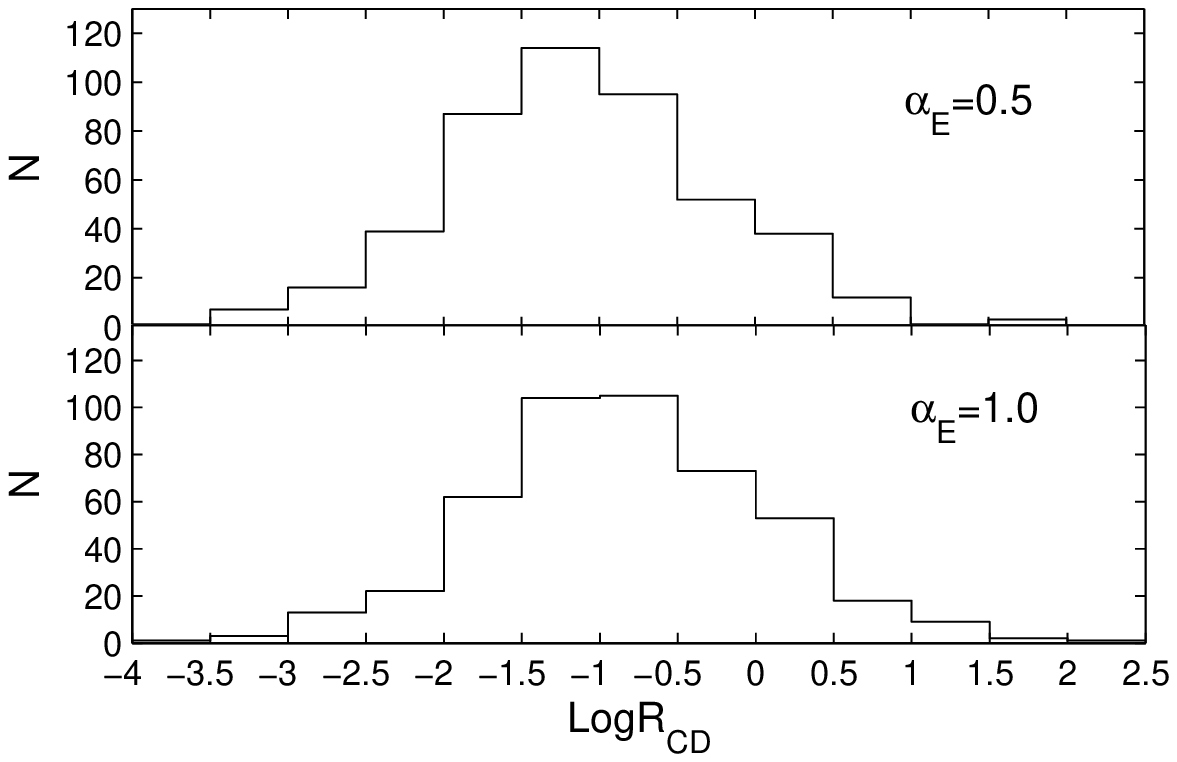}
\caption{The log$R_{CD}$ histogram for FAGNs and NFAGNs according to different extended spectral index $\alpha_{E}$. The top and bottom panels are the cases for FAGNs and NFAGNs respectively, of which the upper panel is corresponding to $\alpha_{E}$=0.5, the bottom panel is $\alpha_{E}$=1.0 respectively.}
\label{R_dis}
\end{figure*}

Fig.\ref{R_dis} shows the $R_{CD}$ distribution of FAGNs and NFAGNs with different extended spectral index, $\alpha_{E}$($\alpha_{E}$=0.5 and $\alpha_{E}$=1.0 respectively). Through the comparison, we can learn that the distribution of $R_{CD}$ from different extended spectral index $\alpha_{E}$ is similar for both FAGNs and NFAGNs. Because of their similarity, we adopt only $\alpha_{E}$=1.0 for the rest of our results. From Fig.1, we can also find that the $R_{CD}$ of NFAGNs are on average smaller than FAGNs for both of $\alpha_{E}$ cases. Using non-parametric Kolmogorov-Smirnov(KS) test, we get that the $R_{CD}$ distribution between FAGNs and NFAGNs are significantly different(chance probability ${\rm{P\sim10^{-17}}}$), the mean values for FAGNs and NFAGNs are Log$R_{CD}$=0.13 and Log$R_{CD}$=-0.86 respectively.

\begin{figure*}
\centering
 \includegraphics[width=8.5cm, height=7.5cm]{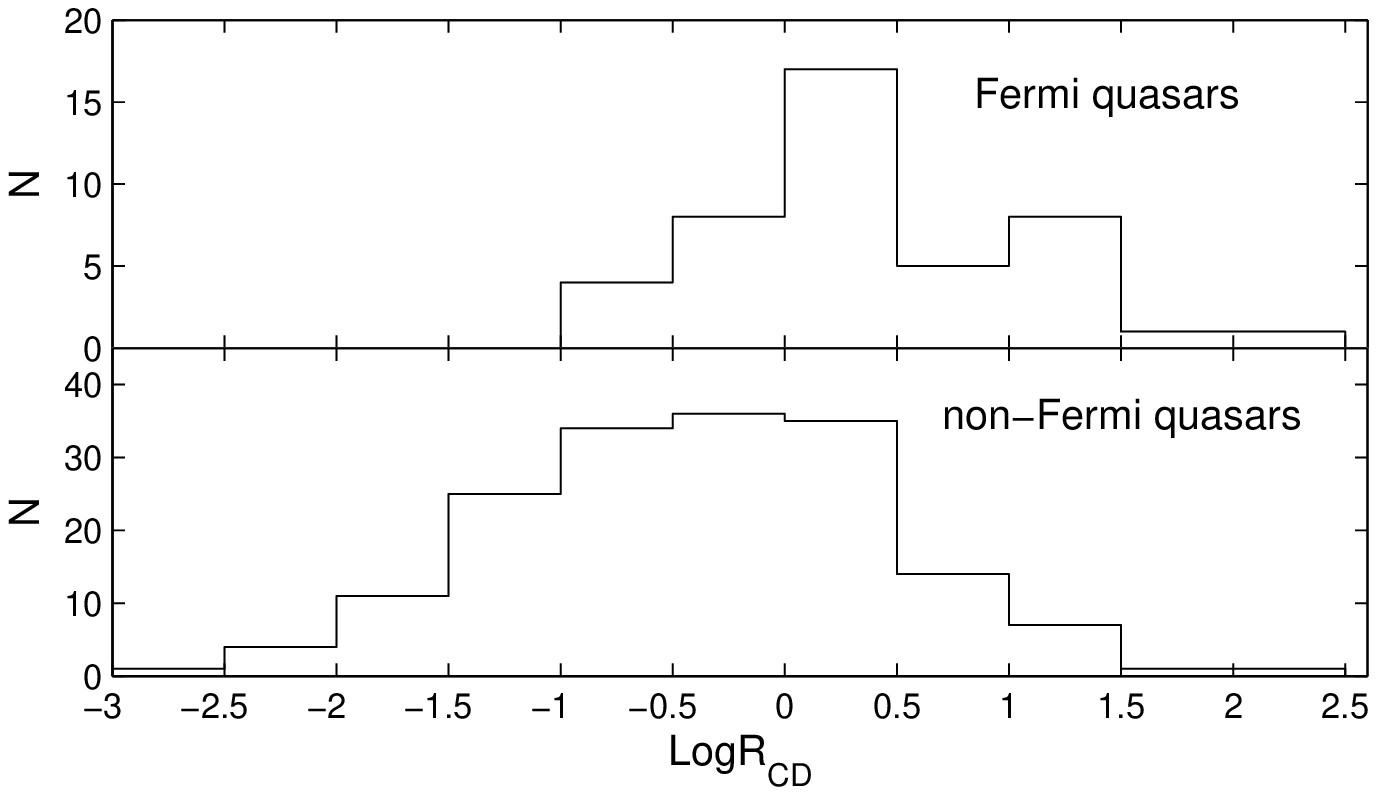}
 \includegraphics[width=8.5cm, height=7.5cm]{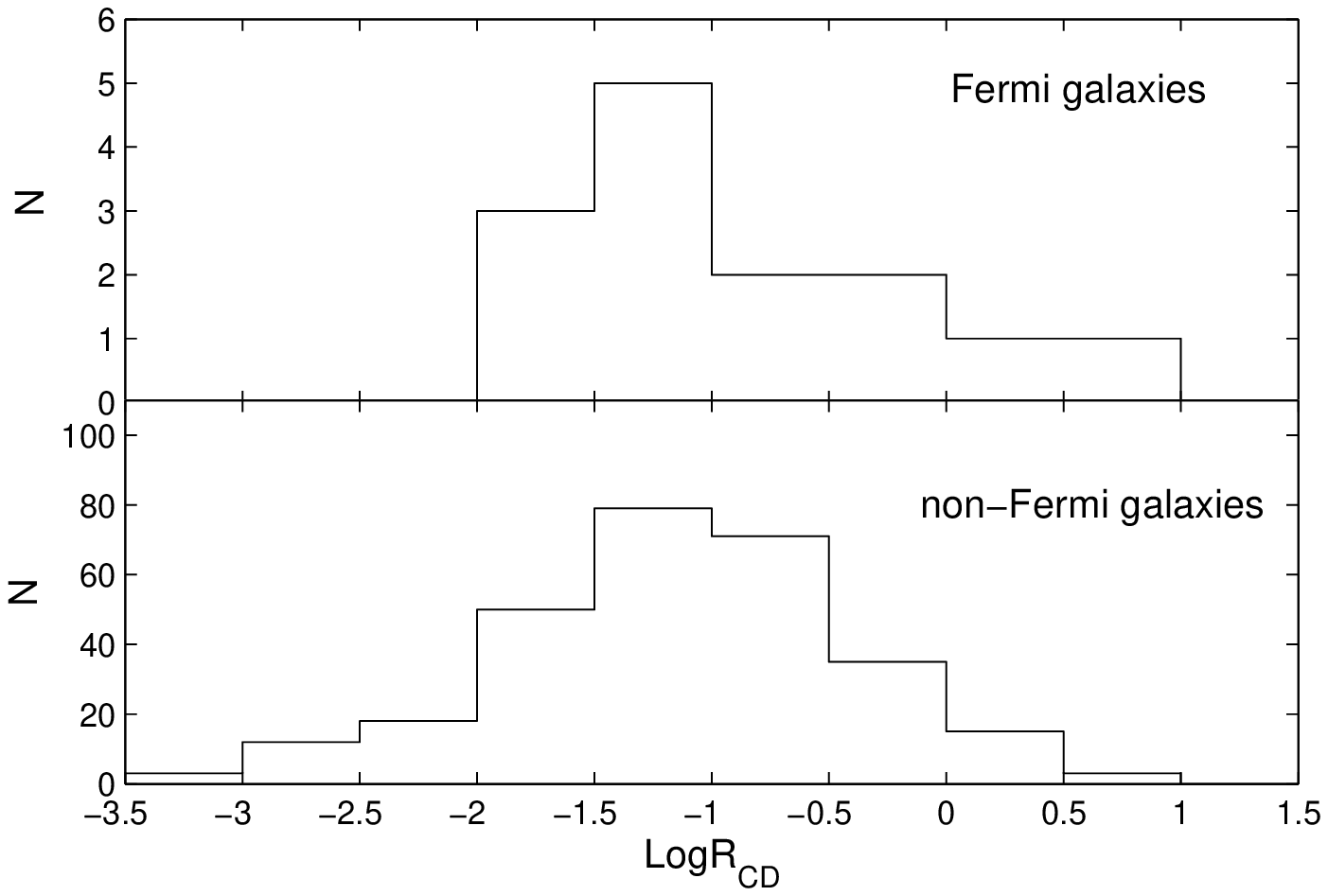}
 \caption{The histogram comparison of $R_{CD}$ for quasars and galaxies(for each picture the upper panel is \emph{Fermi} sources and bottom panel is non-\emph{Fermi} sources).}
 \label{R_hist}
\end{figure*}

The $R_{CD}$ distribution of quasars and radio galaxies are shown in Fig.\ref{R_hist}.  Through nonparametric KS test, we find that the $R_{CD}$ distribution between the \emph{Fermi} quasars and non-\emph{Fermi} quasars are significantly different(chance probability $P\thicksim10^{-7}$), the mean values are Log$R_{CD}$=0.40 and Log$R_{CD}$=-0.43 respectively. While between the \emph{Fermi} galaxies and non-\emph{Fermi} galaxies, the result of KS test shows that there is no significant difference(P=0.713), but the mean value of $R_{CD}$ for the \emph{Fermi} galaxies(Log$R_{CD}$=-0.89) is also higher than the value of non-\emph{Fermi} galaxies(log$R_{CD}$=-1.14). Because the number of \emph{Fermi} galaxies are very small, which is only 14 in the 300 galaxies, this result might not definite. Additionally, because majority of the BL Lacs(22 of 27) in this sample are detected by \emph{Fermi}, the difference of FBL and NFBLs is not studied in this work, the result of BL Lacs is referred to \cite{wu14}.

\subsection{The radio emission of FAGNs and NFAGNs}

\begin{figure}[htb]
\centering
 \includegraphics[width=14cm,height=16cm]{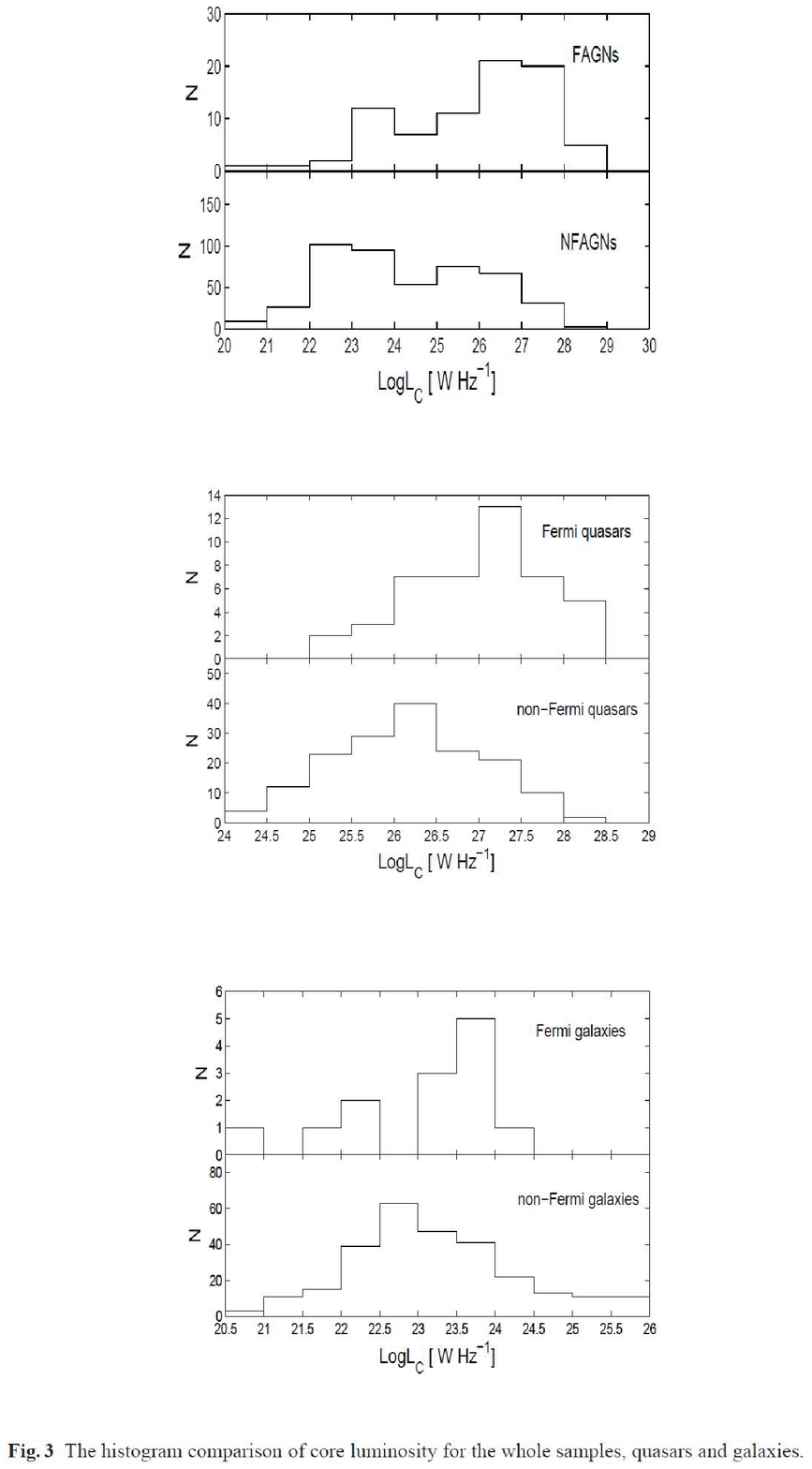 }
 \label{Lcor}
\end{figure}

In this part, we studied the difference of radio core luminosity for \emph{Fermi} and non-\emph{Fermi} sources and shown in Fig.\ref{Lcor}. It shows a tendency that the sources detected with \emph{Fermi} have on average higher core-luminosity than sources not detected. From KS test, the distribution of core luminosity between FAGNs and NFAGNs are significant different ($P\thicksim10^{-9}$ for all,$P\thicksim10^{-5}$ for quasars only).  While for galaxies, the result of KS test shows that they do not have significant differences(P=0.458), but the mean values for \emph{Fermi}-galaxies(Log $L_{c}$=23.29) is slightly higher than the value of  non-\emph{Fermi} galaxies (Log $L_{c} $=23.15).

\begin{figure}
\centering
 \includegraphics[width=14cm]{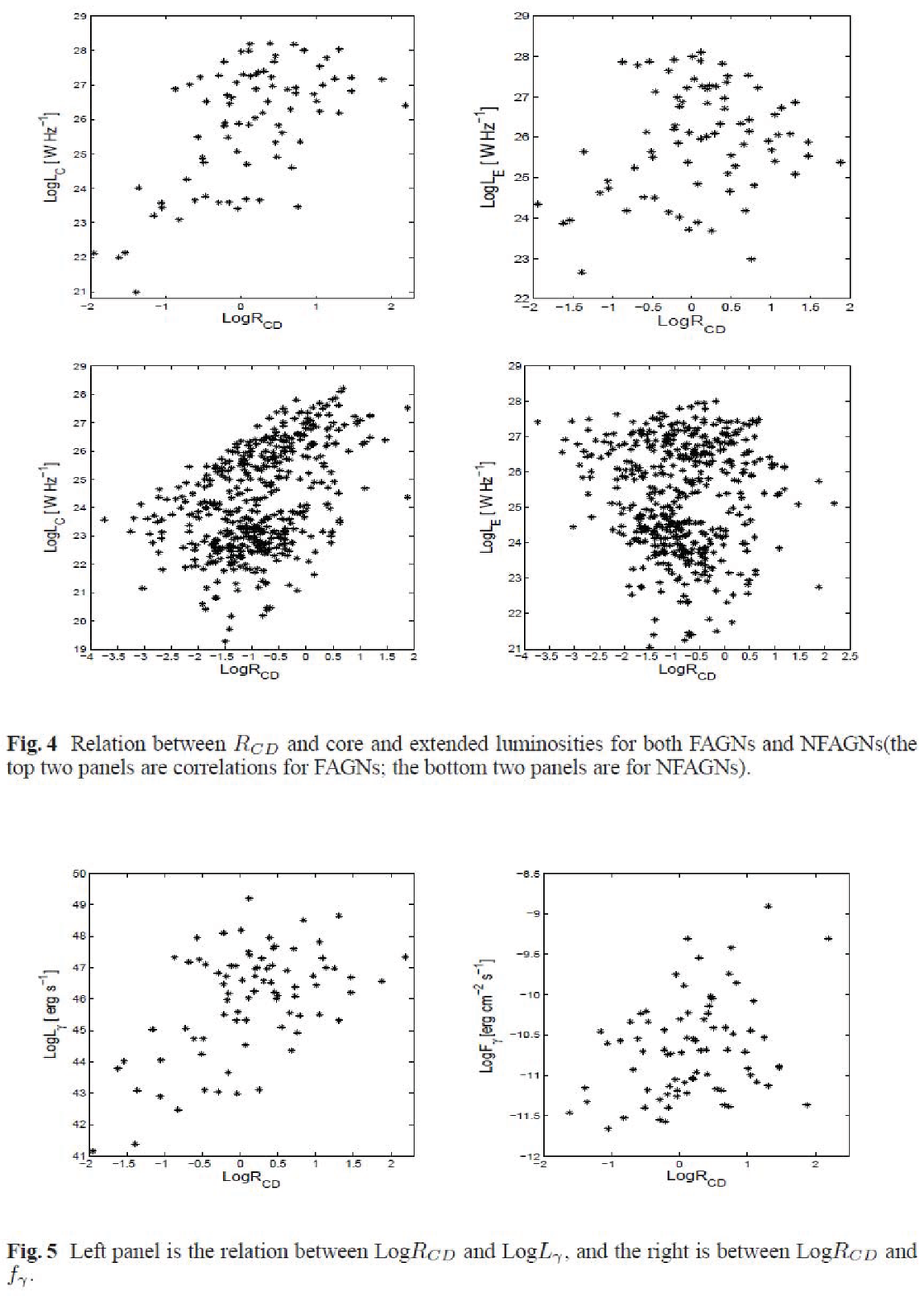 }

 \label{R_corr}
\end{figure}

The relations between $R_{CD}$ and log$L_{C}$, and between $R_{CD}$ and log$L_E$ are all studied and shown in Fig. \ref{R_corr}. We find there are positive correlation between $R_{CD}$ and $L_c$ for both FAGNs and NFAGNs, with correlation coefficient r=0.49 and 0.46 respectively, and all at a confidence level over 99.99$\%$ by Spearman rank correlation analysis. According to the beaming model of AGN\citep{urry95}, the parameters $R_{CD}$ and $L_{C}$ both rely on Doppler factor $\delta$, $R_{CD}$= $R'_{CD}$ $\times$ $\delta^2$ and $L_{C}$ = $L'_{CD}$ $\times$ $\delta^2$(assume the spectral index $\alpha$=0), where $R'_{CD}$ and $L'_{CD}$ are intrinsic core dominance parameter and core luminosity respectively. Because radio galaxies are believed to be the parent population of blazars, \cite{fan03} and this work found that there is no correlation between $R_{CD}$ and $L_{C}$ for radio galaxies, which indicate that the $R'_{CD}$ and $L'_{CD}$ is probably not related. This suggest that the strong correlation between $R_{CD}$ and log$L_{C}$ is probably because they are both relied on $\delta$, that means beaming play a import role in the detected radio core flux and $R_{CD}$ is good indicator of Doppler factor, this is consistent with the beaming model of AGN \citep{urry95}. Meanwhile, for $R_{CD}$ and $L_{E}$, there is no significant correlation; we also found there is no significant differences for the distribution of $L_{E}$ between \emph{Fermi}-QSOs and non-\emph{Fermi}-QSOs, between Fermi-galaxie and non-Fermi galaxies, which indicate that the extend luminosity is less influenced by the beaming effect.

Because $\delta$ is important parameter for the detection of radio flux,  the systematically higher mean and median radio core luminosity in FAGNs indicates that the $gamma$-ray detection of FAGNs might be caused by their higher beaming effect, while we can not exclude the possibility that their intrinsic flux might also plays a role.

\subsection{The $\gamma$-ray emission and $R_{CD}$}

\begin{figure}
\centering
 \includegraphics[width=14cm]{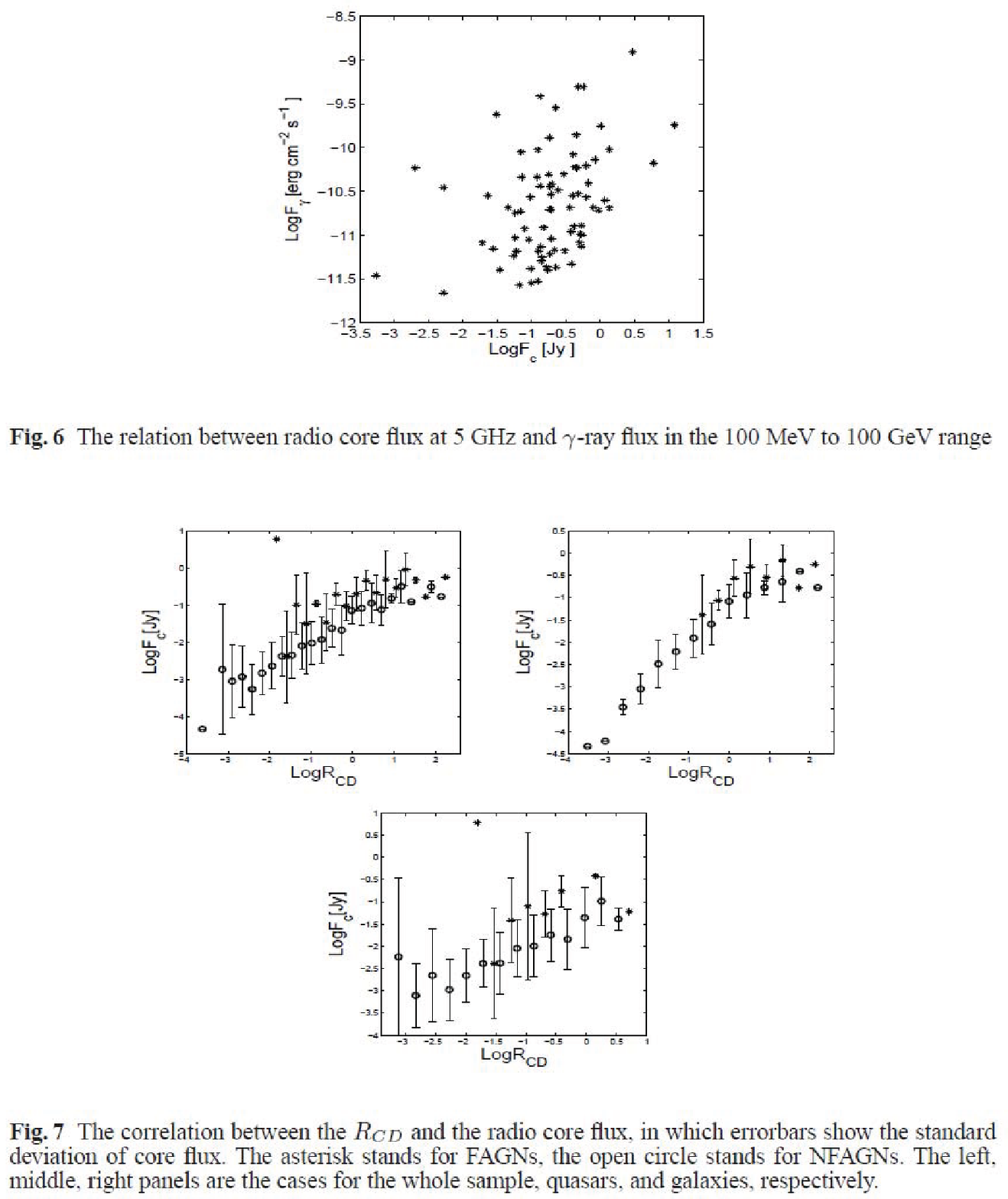 }

 \label{R_Landz}
\end{figure}

 We have obtained the $\gamma$-ray flux in the 100 MeV to 300 GeV energy range for 80 sources in \cite{fan03} from the 3FGL, and calculated the $\gamma$-ray luminosity.  We found a strong correlation between $R_{CD}$ and $L_{\gamma}$ , with a correlation coefficient r=0.39 at $>$ 99.9 percent confidence level, which is shown in the left panel of Fig.5. In addition, we also consider the correlation between $R_{CD}$ and $\gamma$-ray flux $F_{\gamma}$, which is shown in the right panel of Fig.\ref{R_Landz}, with a correlation coefficient r=0.28 at $>$ 98 percent confidence level. Because $R_{CD}$= $R'_{CD}$$\times$ $\delta^2$, these correlation might be caused by the parameters($R_{CD}$ $L_{\gamma}$ and $F_{\gamma}$) all rely on Doppler factor $\delta$ .  These results indicate that $\gamma$-ray emission is probably influenced by jet beaming effect, and $R_{CD}$ can be treated as an indicator of the beaming effect.

\section{Discussion}

In this work, based on a large sample of radio sources with $R_{CD}$ \citep[e.g.,][]{fan03}, we found significant differences in $R_{CD}$ for FAGNs and NFAGNs, \emph{Fermi} quasars and non-\emph{Fermi} quasars.  There is a tendency that the \emph{Fermi} sources have on average higher $R_{CD}$ than the non-\emph{Fermi} sources.  The radio core luminosity of FAGNs are also systematically higher than NFAGNs. These results suggest that \emph{Fermi} sources are probably with strong beaming effect, consistent with results in the literature(eg. \cite{wu14},\cite{2015MNRAS.451.4193C}) and indicate that $R_{CD}$ is probably an indicator of jet beaming effect and  plays an important role in the $\gamma$-ray detection among AGNs in this present sample.

\subsection{The correlation between radio core flux and $\gamma$-ray emission}

\cite{Ghirlanda2011}, \cite{2011ApJ...741...30A} and \cite{ackermann15} all show that there is a statistical significant positive correlation between the centimeter radio and the $\gamma$-ray energy flux.  \cite{wu14} show a significant correlation between $\gamma$-ray flux and radio core flux for a sample of BL Lac objects, a similar correlation is also found for our present sample, see Fig. \ref{fcandgam}. Because $\gamma$-ray flux and radio core flux are Doppler boosted, a strong correlation between them is expected, after excluding
the common dependence on the $R_{CD}$ which is a indicator of Doppler factor by using
the partial Spearman correlation method with a correlation coefficient of 0.33 at $>$ 99 percent level.
Considering the correlation between radio core flux $F_{c}$ and $\gamma$-ray flux $F_\gamma$, NFAGNs maybe have both smaller $F_{c}$ and smaller $F_\gamma$, even though they have comparable $R_{CD}$ with FAGNs. Which makes them more difficult to be detected by \emph{Fermi}-LAT.

\subsection{Why are some sources detected with \emph{Fermi}, but others not?}
\cite{wu14} indicate that the Doppler factor is an important parameter of
$\gamma$-ray detection, the non-detection of $\gamma$-ray emission in NFBLs is likely due to low beaming effect, and/or low intrinsic $\gamma$-ray
flux.
The one important aim of this paper is to test if the results for BL Lac sample in \cite{wu14} is still valid for other type of AGN sample.
We studied the differences of FAGNs and NFAGNs through radio core flux at fixed $R_{CD}$. In Fig.7, we show the correlation between $R_{CD}$ and the average $F_{c}$ of FAGNs and NFAGNs in $R_{CD}$ bins similar with\cite{wu14}. The panels from left to right indicate the cases for the whole sample, quasars and galaxies, respectively (with bin size of 0.24, 0.4 and 0.28 for Log R$_{CD}$). From these figures, we can see that FAGNs have systematically larger radio core flux than NFAGNs at fixed $R_{CD}$, indicating larger intrinsic radio core flux in FAGNs, this result is in consistent with the result of BL Lac objects in \cite{wu14}.

Because FAGNs have systematically larger radio core flux than NFAGNs at fixed $R_{CD}$, then their extended flux are also expected to be larger. Considering the strong linear correlation between intrinsic radio core emission and extend emission \citep{2001ApJ...552..508G}, the extended flux for FAGNs should be also larger than NFAGNs because of their systematically larger radio core flux,  but no strong correlations found between extend emission and $\gamma$-ray emission for this sample. This maybe caused by our sample is small and the result indicate that the intrinsic emission are one possible factor but might not be the crucial factor for the detection of $\gamma$-ray emission as Doppler factor, further study of a larger sample of $\gamma$-ray AGNs might find the correlations between extend radio emission and $\gamma$-ray emission and test our predications.

Together with the results in \cite{wu14}, we can see that \emph{Fermi} detected BL Lacs, QSOs and radio galaxies all have larger intrinsic radio core flux than their non-detected samples. These results indicate a strong connection between radio and $\gamma$-ray emission for the present sample, and it seems to be in favour of the far-dissipation scenario present by \cite{2015MNRAS.452.1280R} and \cite{2011A&A...535A..69N}: a co-spatial origin of the activity for the radio and $\gamma$-ray emission, suggesting that the origin of the seed photons for the high-energy $\gamma$-ray emission is within the jet.

\section{Summary}

In this paper,  we have compared the multiple parameters for FAGNs and NFAGNs by using the available data from the literature. We found that $R_{CD}$ have clear correlations with core luminosity, $\gamma$-ray luminosity, and $\gamma$-ray flux . The average $R_{CD}$ in the \emph{Fermi} sources is on average larger than that in the non-\emph{Fermi} sources.  Moreover, there is a tendency that the \emph{Fermi} sources have higher core-luminosity than the non-\emph{Fermi} sources for the whole sample, quasars and galaxies respectively. We also show that FAGNs have systematically larger radio core flux than NFAGNs at fixed $R_{CD}$, indicating larger intrinsic radio core flux in FAGNs.

Our results indicate that the $R_{CD}$ is an important role of jet beaming effect in $\gamma$-ray detection and show that the beaming effect is vital for the detection of $\gamma$-ray emission. The non-\emph{Fermi} sources is likely due to low beaming effect, and/or the low intrinsic $\gamma$-ray flux. The strong connections between radio and $\gamma$-ray emission might suggest that the origin of the seed photons for the high-energy $\gamma$-ray emission is within the jet for this AGN sample. On account of our sample is limited by the available archival data, the future larger sample of new observational data including redshift, radio core luminosity, extended luminosity and $\gamma$-ray luminosity will be used for further tests of our results.



\begin{table}[t]
\begin{center}
\caption[]{The various parameters for $\gamma$-ray detected sources from \cite{fan03}.}
\label{table:statistics1}
\tiny
  \begin{tabular}{llp{0.5cm}cccccccccc}
  \hline\noalign{\smallskip}
  $Name$ & $ID$ & $z$  & $log {L_{T}}$  &  $log {L_{C}}$ &  $log R_{CD}$ & $log R_{CD}$ & $L_\gamma$   \\
         &     &      &  W Hz$^{-1}$   &W Hz$^{-1}$   &$\alpha_{E}$
=0.5& $\alpha_{E}$
=1.0  &  erg s$^{-1}$    \\
  \hline\noalign{\smallskip}

0414+009 &BL &0.287 &25.30 &24.70 &-0.20 &0.08     &   45.33        \\
0521-365 &BL &0.061 &25.83 &24.75 &-0.77 &-0.49    &   44.74          \\
0548-322 &BL &0.069 &24.39 &23.60 &-0.44 &-0.16    &   43.66          \\
0723-008 &BL &0.130 &25.99 &24.89 &-0.79 &-0.51    &   44.25         \\
0828+493 &BL &0.548 &26.73 &25.90 &-0.48 &-0.21    &   45.51          \\
0829+046 &BL &0.180 &25.55 &25.35 &0.51 & 0.79     &    45.47     \\
0954+658 &BL &0.386 &26.28 &25.48 &-0.45& -0.17    &   45.97         \\
1011+496 &BL &0.200 &25.25 &24.91 &0.20 &0.48      &   46.01       \\
1101+384 &BL &0.031 &23.68 &23.47 &0.48 &0.76      &   44.93           \\
1156+295 &BL &0.729 &27.10 &26.99 &0.82 &1.09      &   47.30         \\
1219+285 &BL &0.100 &25.56 &24.26 &-1.00 &-0.72    &   45.07         \\
1413+135 &BL &0.249 &25.91 &25.61 &0.28  &0.55     &    45.11        \\
1652+398 &BL &0.034 &24.30 &23.69 &-0.21 &0.07     &   44.54        \\
1749+096 &BL &0.322 &26.16 &25.83 &0.22  &0.50     &    46.12      \\
1749+701 &BL &0.770 &27.57 &26.52 &-0.73 &-0.46    &   47.10         \\
1803+784 &BL &0.680 &27.34 &26.97 &0.15  &0.42     &    47.08       \\
1807+698 &BL &0.050 &24.84 &24.60 &0.41  &0.68     &    44.36      \\
1826+796 &BL &0.664 &27.39 &26.88 &-0.07 &0.20     &   46.73            \\
2131-021 &BL &0.557 &26.87 &26.82 &1.19  &1.47     &    46.21           \\
2200+420 &BL &0.069 &25.77 &25.07 &-0.33 &-0.05    &   45.31         \\
2201+044 &BL &0.028 &24.10 &23.41 &-0.31 &-0.04    &   42.99         \\
2240-260 &BL &0.774 &26.87 &26.73 &0.70  &0.97     &    46.74       \\
0305+039 &G &0.029 &24.83 &23.77 &-0.74 &-0.47     &   43.11          \\
0518-458 &G &0.034 &25.93 &24.01 &-1.64 &-1.36     &   43.10         \\
0755+379 &G &0.041 &24.49 &23.59 &-0.57 &-0.29     &   43.05         \\
0909+162 &G &0.085 &24.17 &21.99 &-1.90 &-1.62     &   43.79        \\
1010+350 &G &1.414 &27.14 &26.87 &0.34  &0.62      &    46.91       \\
1253-055 &G &0.014 &24.23 &22.13 &-1.82 &-1.54     &   44.02        \\
1322-427 &G &0.001 &24.62 &22.12 &-2.22 &-1.95     &   41.16          \\
1343-601 &G &0.012 &25.20 &23.58 &-1.33 &-1.06     &   42.91           \\
1441+522 &G &0.140 &25.05 &23.44 &-1.32 &-1.05     &   44.06     \\
1641+399 &G &0.110 &24.93 &23.21 &-1.44 &-1.16     &   45.04           \\
1823+568 &G &0.088 &24.84 &23.65 &-0.88 &-0.61     &   44.74            \\
1142+198 &S &0.021 &24.48 &23.09 &-1.10 &-0.82     &  42.47             \\
0240-002  &S1 &0.004 &22.94 &20.99 &-1.67 &-1.39   &   41.39           \\
1637+826  &S2 &0.023 &24.14 &23.66 &-0.03 &0.25    &   43.12         \\
0202+149 &Q &0.833 &27.61 &27.23 &0.13    &0.41    &    46.53          \\
0212+735 &Q &2.367 &28.59 &28.20 &0.11    &0.39    &   47.95           \\
0333+321 &Q &1.258 &28.36 &27.23 &-0.82&-0.54      &  47.26          \\
0420-014 &Q &0.915 &27.82 &27.26 &-0.14&0.13       &  47.40         \\
0528+134 &Q &2.070 &28.62 &27.97 &-0.26&0.01       &  48.19           \\
0605-085 &Q &0.870 &27.86 &27.37 &-0.04&0.23       &  47.00            \\
0637-752 &Q &0.654 &27.84 &27.40 &0.03 &0.31       &   46.58       \\
0707+476 &Q &1.310 &27.81 &27.30 &-0.07&0.20       &  46.96         \\
0745+241 &Q &0.410 &26.56 &25.88 &-0.30&-0.03      &  45.59         \\
0748+126 &Q &0.889 &27.26 &27.21 &1.19 &1.47       &   46.69         \\
0836+710 &Q &2.160 &28.51 &27.67 &-0.50&-0.22      &  48.10          \\
0838+133 &Q &0.684 &27.23 &26.45 &-0.42&-0.15      &  46.18          \\
0859+470 &Q &1.462 &28.17 &27.27 &-0.57&-0.29      &  46.83          \\
0953+254 &Q &0.712 &26.66 &26.53 &0.73 &1.01       &   46.44         \\
1015+359 &Q &1.226 &27.18 &27.16 &1.60 &1.88       &   46.57          \\
1020+400 &Q &1.254 &27.51 &26.70 &-0.46&-0.18      &  46.73          \\
1150+497 &Q &0.334 &26.43 &25.85 &-0.17&0.11       &  46.03          \\
1217+023 &Q &0.240 &25.68 &25.33 &0.18 &0.46       &  46.21           \\
1222+216 &Q &0.435 &26.64 &26.19 &0.02 &0.29       &   47.29          \\
1226+023 &Q &0.158 &27.14 &26.92 &0.46 &0.73       &   46.09          \\
1315+346 &Q &1.050 &26.98 &26.76 &0.46 &0.73       &   46.39       \\
1418+546 &Q &1.440 &28.27 &27.01 &-0.96&-0.68      &  47.18          \\
1451-375 &Q &0.314 &26.36 &26.24 &0.77 &1.05       &   45.51         \\
1508-055 &Q &1.180 &28.32 &26.88 &-1.15&-0.87      &  47.33         \\
1510-089 &Q &0.361 &26.41 &26.40 &1.91 &2.19       &   47.34        \\
1510-089 &Q &2.100 &28.76 &28.19 &-0.16&0.12       &  49.21           \\
1514-241 &Q &1.546 &26.64 &25.49 &-0.84&-0.57      &  47.95          \\
1532+016 &Q &1.440 &27.78 &27.07 &-0.34&-0.06      &  47.06           \\
1611+343 &Q &1.401 &27.88 &27.78 &0.86 &1.14       &   46.99         \\
1622-297 &Q &0.815 &27.26 &27.18 &0.97 &1.25       &   46.96         \\
1624+416 &Q &2.550 &28.57 &27.99 &-0.17&0.11       &  47.50           \\
1633+382 &Q &1.814 &28.19 &28.01 &0.57 &0.84       &   48.50          \\
1638+398 &Q &1.666 &27.66 &27.54 &0.77 &1.05       &   47.82      \\
1800+440 &Q &0.663 &26.56 &26.04 &-0.09&0.19       &  46.25          \\
1828+487 &Q &0.692 &27.94 &27.30 &-0.25&0.03       &  46.61           \\
1842+681 &Q &0.475 &26.54 &26.29 &0.39 &0.66       &   45.56             \\
1849+670 &Q &0.657 &26.93 &26.52 &0.08 &0.36       &   46.96            \\
2007+777 &Q &0.589 &26.65 &25.81 &-0.50&-0.22      &  46.47             \\
2037+511 &Q &1.686 &28.41 &28.18 &0.43 &0.71       &   47.60           \\
2145+067 &Q &0.990 &28.19 &27.84 &0.18 &0.46       &   47.69             \\
2201+315 &Q &0.298 &26.25 &26.18 &1.03 &1.31       &   45.33          \\
2230+114 &Q &1.037 &28.04 &27.68 &0.17 &0.44       &   47.62           \\
2251+158 &Q &0.859 &28.10 &28.03 &1.03 &1.31       &   48.65         \\
2335-027 &Q &1.072 &27.39 &26.63 &-0.40&-0.12      &  47.06         \\

\noalign{\smallskip}\hline
\end{tabular}
\end{center}
\end{table}

\normalem
\begin{acknowledgements}
 We thank the anonymous referee for insightful comments
and constructive suggestions. This work is supported by the National Nature Science Foundation of China (No. U1431111, 11163002, 11473054, U1531245) and  by the Science
and Technology Commission of Shanghai Municipality (grant
14ZR1447100)

\end{acknowledgements}


\end{document}